%% file: byker+niesler_elephant-doa.tex
\documentclass{article}

\usepackage[nonatbib]{arxiv}

\usepackage{graphics} 
\usepackage{graphicx}
\usepackage{mathptmx} 
\usepackage{times} 
\usepackage{amsmath} 
\usepackage{amssymb}  
\usepackage{hyphenat}
\usepackage{siunitx}
\usepackage{tikz}

\usepackage[inline]{enumitem}

\usepackage[acronym]{glossaries}
\input{acronyms}

\newcommand{\phiset}{\ensuremath{\left\{\phi_1, \phi_2, \ldots, \phi_N\right\}}}
\newcommand{\phiseti}{\ensuremath{\left\{\phi_{1,i}, \phi_{2,i}, \ldots, \phi_{N,i}\right\}}}
\DeclareMathOperator*{\argmax}{arg\,max}
\DeclareMathOperator*{\argmin}{arg\,min}

\title{\LARGE \bf Direction of Arrival Estimation of Wide-band Signals with Planar Microphone Arrays}
\author{Rudolf W Byker and Thomas R Niesler \\
Department of Electrical and Electronic Engineering, Stellenbosch University, South Africa}

\begin{document}

\maketitle
\thispagestyle{empty}
\pagestyle{empty}

\begin{abstract}

An approach to the estimation of the \gls{doa} of wide\hyp band signals with a planar microphone array is presented.
Our algorithm estimates an unambiguous \gls{doa} using a single planar array in which the microphones are placed fairly close together and the sound source is expected to be in the far field.
The algorithm uses the ambiguous \gls{doa} estimates obtained from microphone pairs in the array to determine an unambiguous \gls{doa} estimate for the array as a whole.
The required pair\hyp wise \glspl{doa} may be calculated using \glspl{tde}, which may in turn be calculated using cross\hyp correlation, making the algorithm suitable for wide\hyp band signals.
No a~priori knowledge of the true \gls{ssl} is required.
Simulations show that the algorithm is robust against noise in the input data.
An average ratio of approximately 3:1 exists between the input \gls{doa} errors and the output \gls{doa} error.
Field tests with a moving sound source provided \gls{doa} estimates with standard deviations between \ang{20.4} and \ang{15.2}.


\end{abstract}
\glsresetall{}

\section{INTRODUCTION}

The \gls{doa} estimate provided by a linear microphone array is always ambiguous, because there is no change in the measured signals when the \gls{ssl} is reflected along the array axis.
Other non\hyp trivial ambiguities may also exist \cite{proukakis1994study}.

A triangulation algorithm which removes these ambiguities when three or more linear microphone arrays are present has been proposed by Ottoy and De Strycker \cite{ottoy2016improved}.
We extend this approach to allow the estimation of an unambiguous \gls{doa} for a single non\hyp linear planar microphone array.
We assume that the microphones comprising the array are closely spaced relative to the expected distance to the sound source (i.e. the sound source is situated in the far field).
In this situation, the \glsdisp{tde}{Time Delay Estimate (TDE)} obtained from any pair of microphones in the array provides an ambiguous \gls{doa} estimate.
With $M$ microphones, there are $N={M \choose 2}$ possible pairs, providing $2^N$ possible \gls{doa} solutions.
We define an error function that can be minimised to determine the best solution, thereby eliminating the \gls{doa} ambiguities of the individual microphone pairs.

Although the underlying mathematical derivations are similar, our algorithm differs from Ottoy and De Strycker's method in the following ways:
\begin{enumerate}
	\item Ottoy and De Strycker estimate a \gls{ssl}, while we estimate a \gls{doa}.
	\item The method presented by Ottoy and De Strycker requires three or more linear arrays spaced far apart, with the \gls{ssl} residing in the far field with respect to the elements within each array, but in the near field with respect to the arrays themselves.
	We use a single non\hyp linear, planar array, where the \gls{ssl} is in the far field with respect to the elements within the array.
	Of course, triangulation may still be performed after our method has been applied to two or more such arrays.
\end{enumerate}
Both methods aim to resolve the \gls{doa} ambiguities inherent to linear arrays.

In \S\ref{sec:background} we review the existing algorithm presented by Ottoy and De Strycker, and in \S\ref{sec:extend} we present our proposed method.
Simulations (\S\ref{sec:simulations}) and tests (\S\ref{sec:tests}) are performed to verify the algorithm.

\section{BACKGROUND}
\label{sec:background}

\subsection{DOA Estimation with a single microphone pair}

\begin{figure}
	\centering
	\includegraphics[scale=.9]{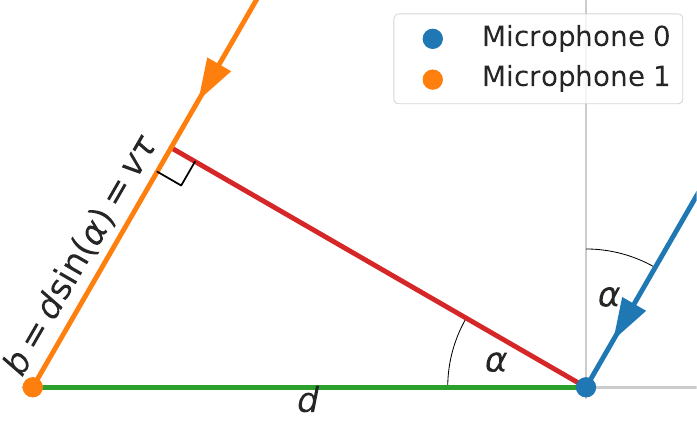}
	\caption{
		When the sound source is in the far field with respect to the microphone pair (i.e. the distance to the sound source is much larger than $d$), the vectors from the \gls{ssl} to the microphones may be considered parallel.
	}
	\label{fig:doa_trigenometrie_alpha}
\end{figure}

\begin{figure}
	\centering
	\includegraphics[scale=.9]{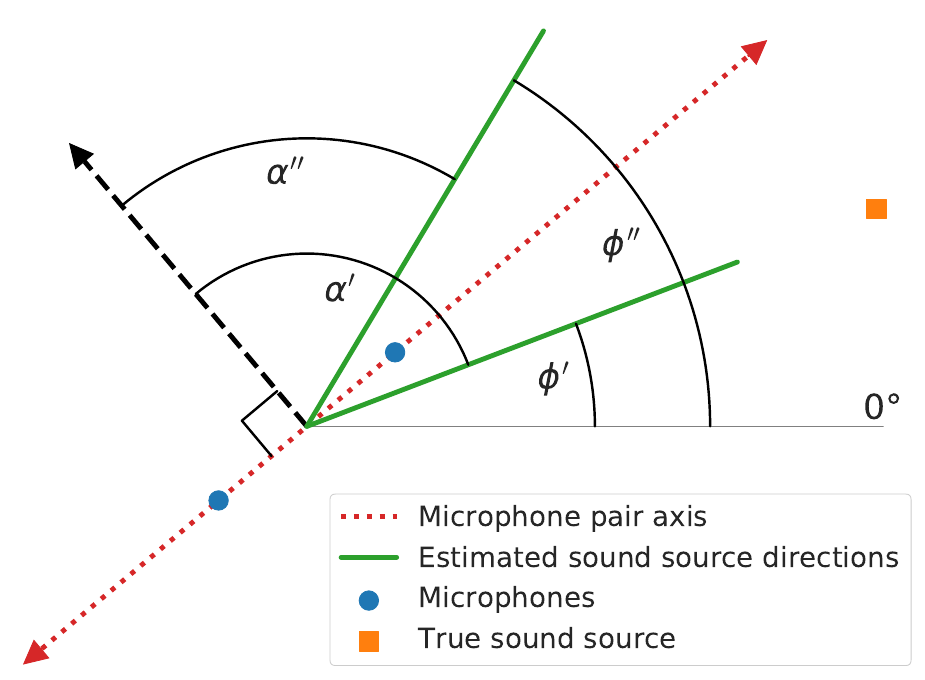}
	\caption{
		Since sound sources with \glspl{doa} $\alpha^\prime$ and $\alpha^{\prime\prime}$ lead to the same \gls{tdoa} for the microphone pair, there is a \gls{doa} ambiguity.
		The \glspl{doa} relative to the coordinate system of the microphone pair are indicated by $\alpha^\prime$ and $\alpha^{\prime\prime}$.
		The \glspl{doa} relative to the polar coordinate system are indicated by $\phi^\prime$ and $\phi^{\prime\prime}$.
	}
	\label{fig:mic_pair_alias_trigenometry}
\end{figure}

When two closely-spaced microphones measure the same far\hyp field signal, the vectors from the \gls{ssl} to each microphone may be regarded as parallel, as shown in Figure~\ref{fig:doa_trigenometrie_alpha}.
The \gls{tdoa} between the microphones relates to the \gls{doa} as indicated by Equation~\ref{eq:pair_doa} where $\alpha$ is the \gls{doa} of the sound source, $v$ is the speed of sound, $d$ is the microphone pair distance, and $\tau$ is the \gls{tdoa} estimate \cite{dostalek2009direction, tashev2009sound}.
The \gls{tdoa} may be estimated using any known method, such as cross correlation.
\begin{equation}\label{eq:pair_doa}
\alpha = \arcsin \left( \dfrac{\tau v}{d} \right)
\end{equation}
With two microphones, the \gls{doa} always has an alias.
As demonstrated in Figure~\ref{fig:mic_pair_alias_trigenometry}, sound sources with \glspl{doa} $\alpha^\prime$ and $\alpha^{\prime\prime}$ have the same \gls{tdoa}, and therefore cannot be distinguished.
The same ambiguity exists in any linear microphone array \cite{proukakis1994study}.

Since each microphone pair has its own coordinate system, the rest of this paper will use the \gls{doa} estimates relative to the polar coordinate system ($\phi$) instead of the \gls{doa} estimates relative to the coördinate systems of the individual microphone pairs ($\alpha$), as illustrated in Figure~\ref{fig:mic_pair_alias_trigenometry}.

\subsection{Triangulation with multiple DOA estimates}
\label{sec:triangulation}

\begin{figure}
	\centering
	\includegraphics[scale=.9]{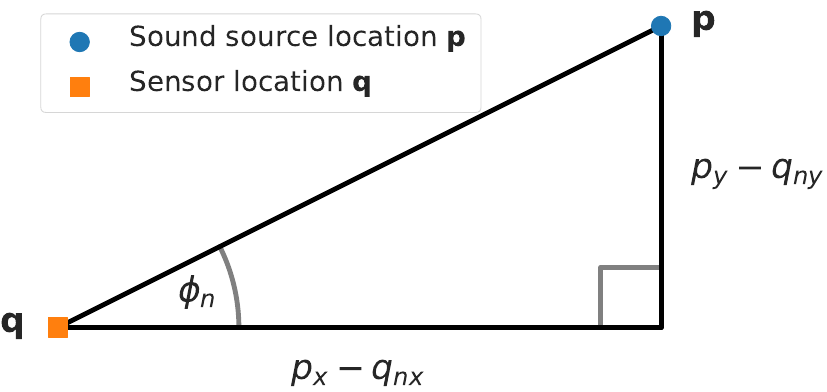}
	\caption{
		The \gls{doa} estimate $\phi_n$ is related trigonometrically to the sensor location $\mathbf{q}_n=[q_{nx}, q_{ny}]$ and the \gls{ssl} $\mathbf{p} = [p_x, p_y]$.
	}
	\label{fig:equation_tan_phi_n}
\end{figure}

When multiple unambiguous \gls{doa} estimates are available from two or more different sensor locations to the same sound source, triangulation may be performed to estimate the \gls{ssl} \cite{ottoy2016improved, hartley1997triangulation}.
A \gls{doa} estimate $\phi_n$, its sensor location $\mathbf{q}_n=[q_{nx}, q_{ny}]$ and the \gls{ssl} $\mathbf{p} = [p_x, p_y]$ are related by Equation~\ref{eq:ottoy_node_tangent} as depicted in Figure~\ref{fig:equation_tan_phi_n}.
\begin{equation}\label{eq:ottoy_node_tangent}
\tan \left( \phi_n \right) = \dfrac{p_y - q_{ny}}{p_x - q_{nx}}
\end{equation}

Since $p_x$ and $p_y$ are unknown, two sensors and two such equations are needed to solve for the \gls{ssl}.
With more than two sensors, an over-determined system of linear equations is produced, as shown in Equation~\ref{eq:ottoy_overdetermined_system}, which may be solved using the least squares method \cite{ottoy2016improved, stigler1981gauss, fuller2009measurement}.
\begin{equation}\label{eq:ottoy_overdetermined_system}
\begin{bmatrix}
-\tan \left( \phi_1 \right) & 1 \\
-\tan \left( \phi_2 \right) & 1 \\
\vdots                     & 1 \\
-\tan \left( \phi_N \right) & 1
\end{bmatrix}
\begin{bmatrix}
p_x \\
p_y
\end{bmatrix}
=
\begin{bmatrix}
q_{1,y} - q_{1,x} \tan \left( \phi_1 \right) \\
q_{2,y} - q_{2,x} \tan \left( \phi_2 \right) \\
\vdots \\
q_{N,y} - q_{N,x} \tan \left( \phi_N \right) \\
\end{bmatrix}
\end{equation}

\subsection{Ottoy and De Strycker's method for solving DOA ambiguities}

An algorithm to overcome the \gls{doa} ambiguity while performing triangulation was previously presented by Ottoy and de Strycker \cite{ottoy2016improved}.
We briefly review this algorithm here.

Each sensor location $\mathbf{q}_n=[q_{nx}, q_{ny}]$ is referred to as a `node'.
This may be a microphone pair, linear microphone array, or any other sensor providing an ambiguous \gls{doa} estimate $\phi_n$.
Each estimate may take on one of two values, which we call $\phi_n^\prime$ and $\phi_n^{\prime\prime}$.
If there are $N$ nodes, then there are $I=2^N$ ways to interpret the set of estimates \phiset{}.
For each interpretation \phiseti{}, with $i \in \left\{1, 2, \ldots, I\right\}$, Equation~\ref{eq:ottoy_overdetermined_system} is solved to obtain an estimated \gls{ssl} $\hat{\mathbf{p}}_i = [\hat p_{x, i}, \hat p_{y, i}]$, after which the angle from $\mathbf{q}_n$ to $\hat{\mathbf{p}}_i$ for each $n$ is given by Equation~\ref{eq:hat_phi_n}.
\begin{equation}\label{eq:hat_phi_n}
\hat{\phi}_{n, i} = \arctan \left( \dfrac{\hat{p}_{y,i} - q_{ny}}{\hat{p}_{x,i} - q_{nx}} \right)
\end{equation}
The absolute error associated with any one interpretation \phiseti{} is defined by Equation~\ref{eq:mean_error}.
This is the sum of the absolute differences between the \gls{doa} estimate given by each array and the angle between that array and the estimated \gls{ssl}.
\begin{equation}\label{eq:mean_error}
\phi_{\text{err}, i} = \sum_{n=1}^{N} \left| \phi_{n, i} - \hat{\phi}_{n, i} \right|
\end{equation}
The interpretation where $i=\argmin_i \left\{ \phi_{err, i} \right\}$ is used to calculate the final \gls{ssl}.

\section{EXTENSION TO A SINGLE PLANAR ARRAY}
\label{sec:extend}

The algorithm presented by Ottoy and De Strycker works well for situations in which:
\begin{enumerate}
	\item the microphones within each array are closely spaced in relation to the expected \glspl{ssl}, so that the sound sources are in the far field in relation to the individual microphones, and
	\item the arrays themselves are spaced far apart, so that the sound sources are in the near field with respect to the arrays.
\end{enumerate}

The algorithm can be extended to perform a \gls{doa} estimation for a single planar array for which the sound source is in the far field, by
\begin{enumerate*}
	\item redefining the $N$ `nodes' to be the possible combinations of microphone pairs within the planar array, and
	\item redefining the error function
\end{enumerate*}.
We will accomplish this in the following paragraphs.

\subsection{Pair-wise DOA estimation}

We now regard each possible pair of microphones in the planar array as a distinct two\hyp element linear array, and refer to this as a `node'.
For an array with $M$ microphones, there are $N = {M \choose 2}$ possible nodes.
As before (see Figure~\ref{fig:mic_pair_alias_trigenometry}), each node provides two possible solutions, $\phi_n^\prime$ and $\phi_n^{\prime\prime}$, and the correct solution is one of $I=2^N$ possible interpretations \phiseti{} of the set \phiset{}, with $i \in \left\{1, 2, \ldots, I\right\}$.

\subsection{Error function}

Assuming that the sound source is in the far field, we may consider all microphone pairs to be at the same position.
Hence we can consider the correct \gls{doa} solution to be the one for which the \gls{doa} estimate of all pairs agree.
To find the solution that is closest to this ideal, a new error function is defined.

The values in each set \phiseti{} are discrete points in a continuous space.
To find the mode of such a dataset, \gls{kde} may be performed.
Since the domain of the data is wrapped, the von~Mises kernel, which is an approximation of the wrapped normal distribution, is appropriate \cite{von1918uber, mardia1972statistics, mardia1975statistics}.
The von~Mises kernel is given by Equation~\ref{eq:vonmises}, where $I_0$ is the modified zero'th order Bessel function, $\mu$ is the mean and $\kappa$ is the concentration, which determines the width of the kernel.
Small values of $\kappa$ correspond to wide kernels, and vice versa.
\begin{equation}\label{eq:vonmises}
f_\text{kernel}\left( \phi | \mu, \kappa \right) = \dfrac{\exp\left(\kappa \cos \left( \phi - \mu \right)\right)}{2 \pi I_0 \left( \kappa \right)}
\end{equation}
The \gls{kde} of interpretation $i$ is given by Equation~\ref{eq:vonmises_kde}.
\begin{equation}\label{eq:vonmises_kde}
g_i\left(\phi | \kappa \right) = \dfrac{1}{N} \sum_{n=1}^{N} f_\text{kernel}\left( \phi | \mu=\phi_{n,i}, \kappa \right)
\end{equation}

We define the `concensus' \gls{doa} estimate for interpretation $i$ as the mode of the \gls{kde}, as expressed by Equation~\ref{eq:vonmises_mode}.
Since all nodes are assumed to be at the same position, the subscript $n$ used in Equations~\ref{eq:hat_phi_n} and \ref{eq:mean_error} is not needed.
\begin{equation}\label{eq:vonmises_mode}
\hat{\phi}_i = \argmax_\phi\left\{ g_i\left(\phi\right) \right\}
\end{equation}

The error of interpretation $i$ is defined by Equation~\ref{eq:vonmises_mode_error}.
This is the sum of the absolute differences between the \gls{doa} estimates of the microphone pairs $\phi_{n,i}$ and the consensus \gls{doa} estimate $\hat{\phi}_i$.
\begin{equation}\label{eq:vonmises_mode_error}
\phi_{\text{err}, i} = \sum_{n=1}^{N} \left| \phi_{n,i} - \hat{\phi}_i \right|
\end{equation}

\subsection{Numerical implementation}

Since only the position of the peak of the \gls{kde} is considered, all factors in Equations~\ref{eq:vonmises} and \ref{eq:vonmises_kde} which are independent of $\phi$ may be ignored.
Consequently, the \gls{kde} may be simplified as shown in Equation~\ref{eq:kde_optimise_and_derivatives}.
The first and second derivatives are also shown.
\begin{subequations}\label{eq:kde_optimise_and_derivatives}
	\begin{align}
	g_i\left( \phi | \kappa \right) &= \sum_{n=1}^{N} \exp\left(\kappa \cos \left( \phi - \phi_{n,i} \right)\right) \label{eq:kde_optimise} \\
	g_i^{\prime}\left( \phi | \kappa \right) &= - \kappa \sum_{n=1}^{N} \sin\left( \phi - \phi_{n,i} \right) \exp\left(\kappa \cos \left( \phi - \phi_{n,i} \right)\right) \label{eq:kde_optimise_jac} \\
	g_i^{\prime\prime}\left( \phi | \kappa \right) &=
	\begin{aligned}[t]
		\kappa \sum_{n=1}^{N} &\exp\left(\kappa \cos \left( \phi - \phi_{n,i} \right)\right) \\
		&\quad \left( \kappa \sin^2\left( \phi - \phi_{n,i} \right) - \cos\left( \phi - \phi_{n,i} \right) \right)
	\end{aligned}\label{eq:kde_optimise_hess}
	\end{align}
\end{subequations}

The peak of $g_i\left( \phi | \kappa \right)$ may be found using any suitable numerical optimisation function.
Since the first and second derivatives are available in closed form and not expensive to calculate, they can be taken advantage of, for example by the \gls{ncg} method \cite{nocedal2006numerical}, which often converges more quickly than other methods.

A different \gls{kde} $g_i\left( \phi | \kappa \right)$ is calculated for each of the $I$ interpretations of \phiseti{}.
The interpretation for which $\phi_{\text{err},i}$ (Equation~\ref{eq:vonmises_mode_error}) is the smallest is regarded as the correct interpretation, and the mode $\hat{\phi}_i$ (Equation~\ref{eq:vonmises_mode}) is the final \gls{doa} estimate.
We will refer to the \gls{kde} of the favoured interpretation as $g_\text{final}\left( \phi | \kappa \right)$, where $\text{`final'} = \argmin_i \left\{ \phi_{\text{err}, i} \right\}$.

\subsection{Initialising the optimisation of $g_i\left( \phi | \kappa \right)$}\label{sec:init}

While $g_i\left( \phi | \kappa \right)$ may be evaluated directly, a brute\hyp force search for the peak is computationally expensive.
This problem is compounded by the fact that there are $I=2^N$ interpretations for which the peak of $g_i\left( \phi | \kappa \right)$ must be found.
The optimisation may be accelerated considerably by choosing an appropriate starting value for $\phi$ before applying the \gls{ncg} method to $g_i\left( \phi | \kappa \right)$.

To find an appropriate starting value for the optimisation of $g_i\left( \phi | \kappa \right)$, we calculate an approximation $\hat g_i\left( \phi | \kappa \right)$ of $g_i\left( \phi | \kappa \right)$.
Then, $\argmax_\phi \hat g_i\left( \phi | \kappa \right)$ is used as the starting point for finding the peak of $g_i\left( \phi | \kappa \right)$.
One strategy for fast estimation of $g_i\left( \phi | \kappa \right)$ is presented here.

\begin{figure}
	\centering
	\includegraphics[scale=.9]{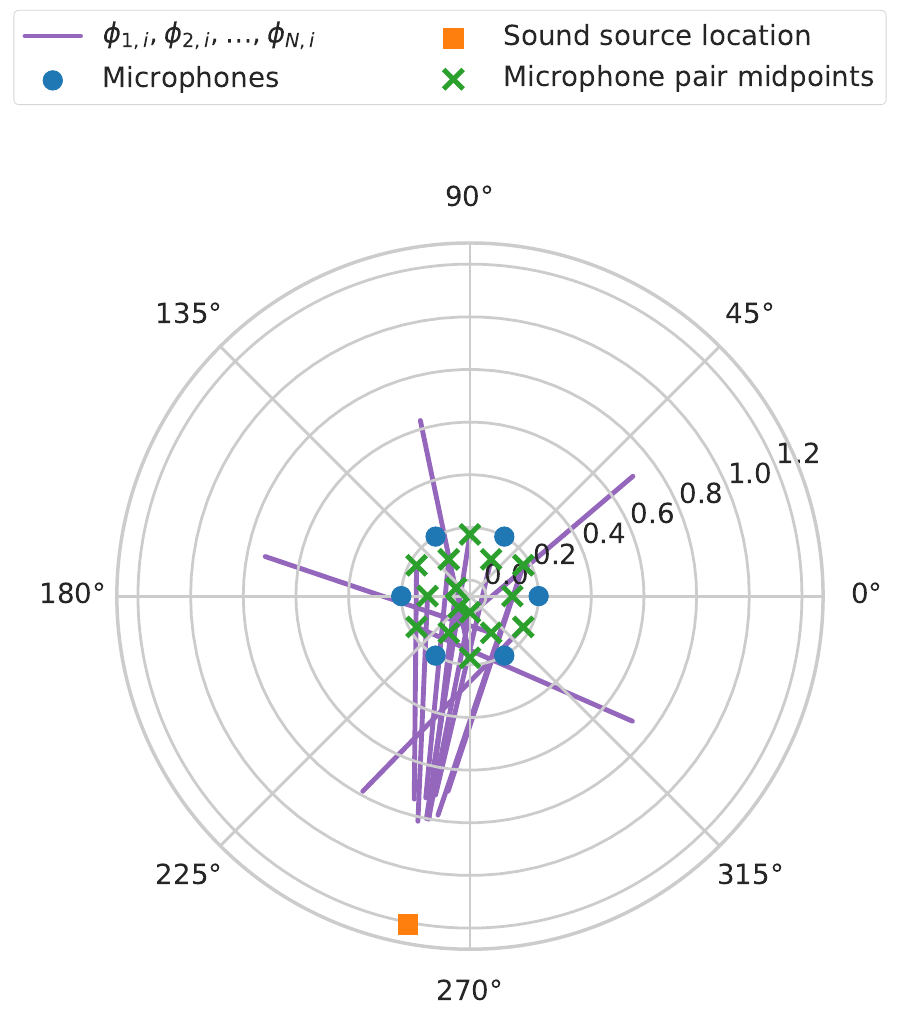}
	\caption{
		An example of a dataset \phiseti{} with $N=15$ and where \num{5} of the values are interpreted incorrectly.
	}
	\label{fig:init_demo_polar}
\end{figure}

\begin{figure}
	\centering
	\includegraphics[scale=.9]{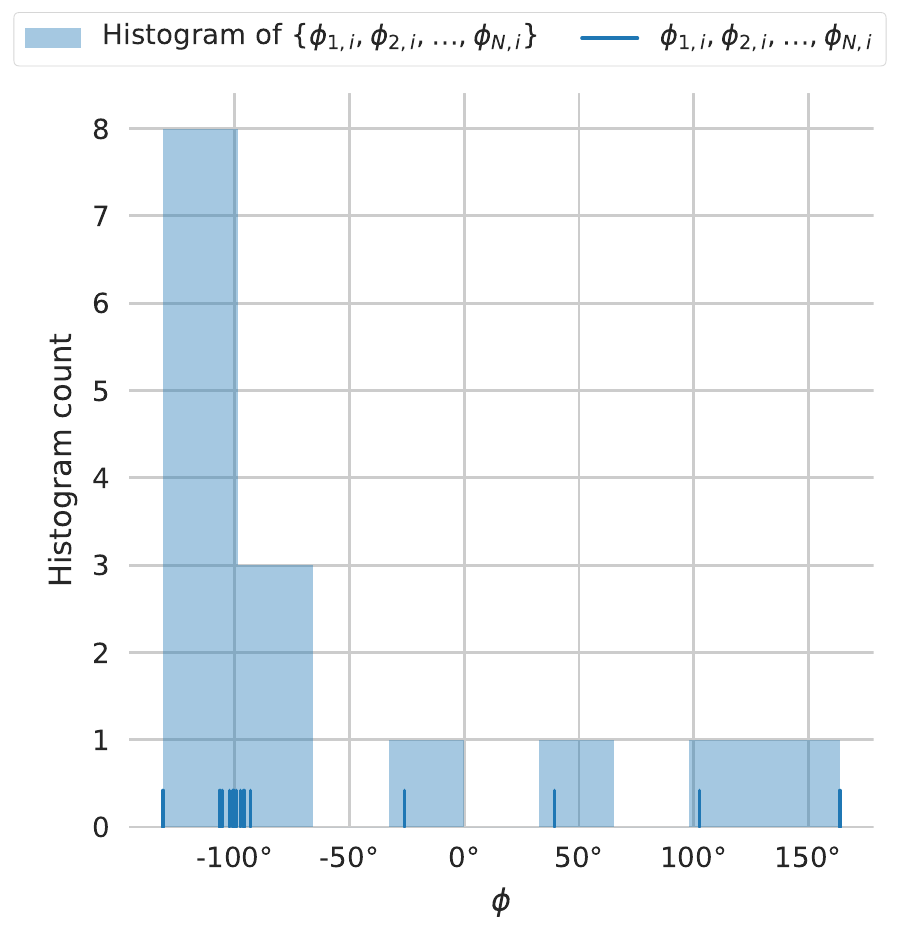}
	\caption{
		A histogram of the dataset from Figure~\ref{fig:init_demo_polar} using 8 bins.
		The width of each bin is \ang{45}.
	}
	\label{fig:init_demo_histogram}
\end{figure}

A histogram of the values in \phiseti{} approximates the true distribution as the number of bins increases.
Therefore, the convolution of such a histogram with the von~Mises kernel approximates $g_i\left( \phi | \kappa \right)$.
An example of a dataset \phiseti{} is shown in Figure~\ref{fig:init_demo_polar}, and Figure~\ref{fig:init_demo_histogram} shows its \num{8}\hyp bin histogram.
This histogram may be convolved with the von~Mises kernel to obtain the approximation $\hat g_i\left( \phi | \kappa \right)$,
after which $\argmax_\phi \hat g_i\left( \phi | \kappa \right)$ is used to initialise the \gls{ncg} optimiser.

The convolution may be implemented using the \gls{fft}, and the \gls{dft} of the von~Mises kernel can be precomputed.
This allows $\hat g_i\left( \phi | \kappa \right)$ to be computed efficiently.
A large number of histogram bins leads to slower calculation of $\hat g_i\left( \phi | \kappa \right)$, but faster optimisation of $g_i\left( \phi | \kappa \right)$.
A smaller number of bins leads to faster calculation of $\hat g_i\left( \phi | \kappa \right)$, but slower optimisation of $g_i\left( \phi | \kappa \right)$.
The best compromise will depend on the exact implementation used and the capabilities of the processors performing the calculations.

The \gls{ncg} optimisation step may be skipped if speed is valued over accuracy, or if the selected number of histogram bins already places $\argmax_\phi \hat g\left( \phi | \kappa \right)$ within the desired accuracy.

\section{Simulations}
\label{sec:simulations}

\begin{figure}
	\centering
	\includegraphics[scale=.9]{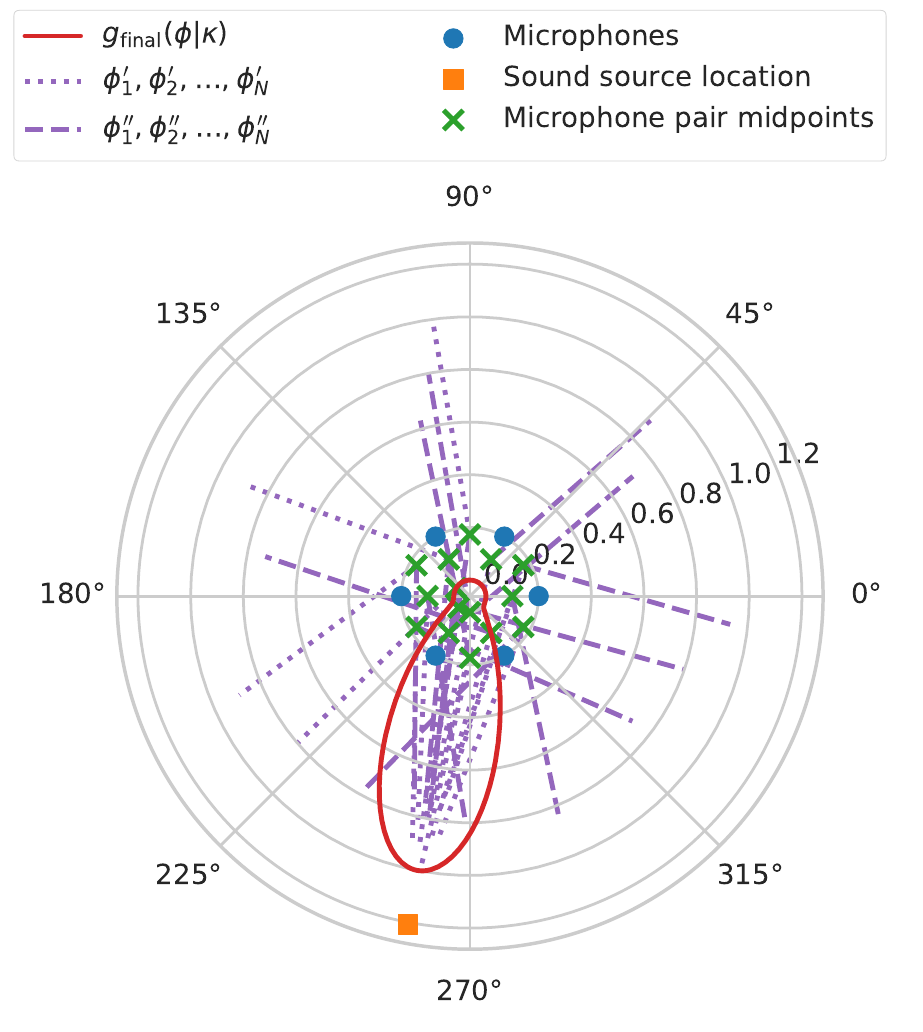}
	\caption{
		This circular array of 6 microphones, with a radius of \SI{20}{\centi\meter} provides 15 ambiguous \gls{doa} estimates.
		Each ambiguous estimate has one interpretation pointing towards the sound source, and one which is mirrored along the microphone pair axis.
		The \gls{kde} $g_\text{final}\left( \phi | \kappa \right)$ yielding the lowest error $\phi_{\text{err}, i}$ is also shown.
		The mode of this \gls{kde} is the \gls{doa} estimate provided by our algorithm.
	}
	\label{fig:sim_demo}
\end{figure}

The presented algorithm was developed as part of a project which sought to localise elephant rumbles using microphone arrays.
Recording devices with circular arrays of \num{6} omnidirectional microphones with a radius of \SI{20}{\centi\meter} were used (see Figure~\ref{fig:sim_demo}).
For this configuration there are $N=\num{15}$ microphone pairs and $I=\num{32768}$ possible interpretations of \phiset{} to consider per microphone array.
One such circular array is used in the following simulation.

The effect of noisy input \gls{doa} estimates on the output \gls{doa} estimate is also considered in the simulation.
The propagation of sound from the \gls{ssl} to the sensors is not simulated.
Therefore, the results are expected to be independent of the \gls{ssl}, but dependent on the level of noise in the input data.

To initialise the optimisation of $g_i\left( \phi | \kappa \right)$, the distribution was first approximated by a \num{512}\hyp bin histogram of \phiseti{} convolved with the von~Mises kernel, as suggested in \S\ref{sec:init}.
This brings the initial value to within \ang{0.703} of the final \gls{doa} prediction.
Starting from this initialisation, a \gls{ncg} optimiser locates the peak of $g_i\left( \phi | \kappa \right)$, .

Every iteration of the simulation consists of the following steps:
\begin{enumerate}
	\item A random \gls{ssl} is chosen in the annulus defined by $\SI{10}{\meter} < r < \SI{1}{\kilo\meter}$, where $r$ is the distance from the origin.
	The origin is also the midpoint of the circular microphone array.
	\item The true \gls{doa} from the midpoint of each microphone pair to the sound source is calculated.
	\item Gaussian random values with zero mean and a standard deviation of $\sigma$ are added to the true \glspl{doa}, where $\sigma$ is selected uniformly from $[\ang{0}; \ang{45})$.
	This represents the estimated \gls{doa} values from each microphone pair which would constitute the inputs to our algorithm during real\hyp world application.
	\item A random value for the concentration $\kappa$ is chosen uniformly from $[\num{.1}; \num{100})$.
	\item Our algorithm is used to estimate a single \gls{doa} for the microphone array.
	\item The estimated \gls{doa} is compared to the true \gls{doa}.
	The algorithm error is defined as the magnitude of the smallest angle between the estimated \gls{doa} and the true \gls{doa}.
\end{enumerate}

A total of \num{e5} simulation iterations were performed.

An example of the \gls{doa} estimates and the \gls{kde} $g_\text{final} \left( \phi | \kappa \right)$ of the interpretation yielding the lowest error with this configuration of microphones is shown in Figure~\ref{fig:sim_demo}.

\begin{figure}
	\centering
	\includegraphics[scale=.9]{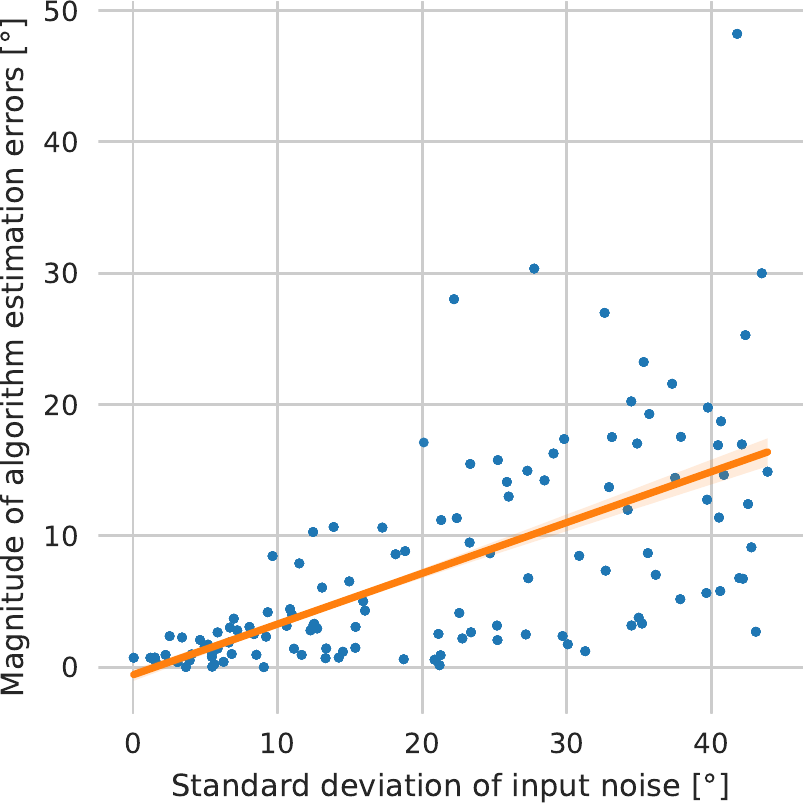}
	\caption{
		The relationship between the standard deviation of the noise added to the input \gls{doa} estimates (step 3 in \S\ref{sec:simulations}) and the algorithm estimation error (defined in step 6 in \S\ref{sec:simulations}).
		Errors in the input data are positively correlated with errors in the estimated \gls{doa}, but the latter is on average only a third of the former.
		Shading around the regression line indicates the 95\% confidence intervals.
		For clean input data, the convolution with the von~Mises kernel and the subsequent optimisation introduces some numerical error.
		In real-world situations, the noise in the input data is expected to be much greater than this numerical noise.
	}
	\label{fig:sim01_doa_gaussian_error}
\end{figure}

Figure~\ref{fig:sim01_doa_gaussian_error} depicts the correlation between the standard deviation of noise in the input data, $\sigma$, and the output error.
On average, a 3:1 reduction in \gls{doa} noise is achieved, indicating that the algorithm is robust against noisy input data.

The polar coordinates of the \gls{ssl} $\left(r; \phi\right)$ showed no significant correlation with the dependent variable.
The lack of correlation between the value of the \gls{ssl} coordinates and the output error was to be expected, since the simulation was set up to ensure that the distance between the sound source and the microphone array has no impact on the level of noise in the input data.
In practical tests, it is expected that the output error will be affected by the distance to the sound source, since the \gls{doa} estimates from the microphone pairs will be less accurate for distant sound sources than for sound sources nearby, causing a positive correlation between the distance to the sound source and the level of noise in the input data.

There was also no discernible correlation between the concentration $\kappa$ of the von~Mises kernel and the dependent variable for the values of $\kappa$ that were tested.
This suggests that $\kappa$ does not need to be fine\hyp tuned for this algorithm to work properly.

\section{Practical tests}
\label{sec:tests}

\begin{figure}
	\centering
	\includegraphics[scale=.9]{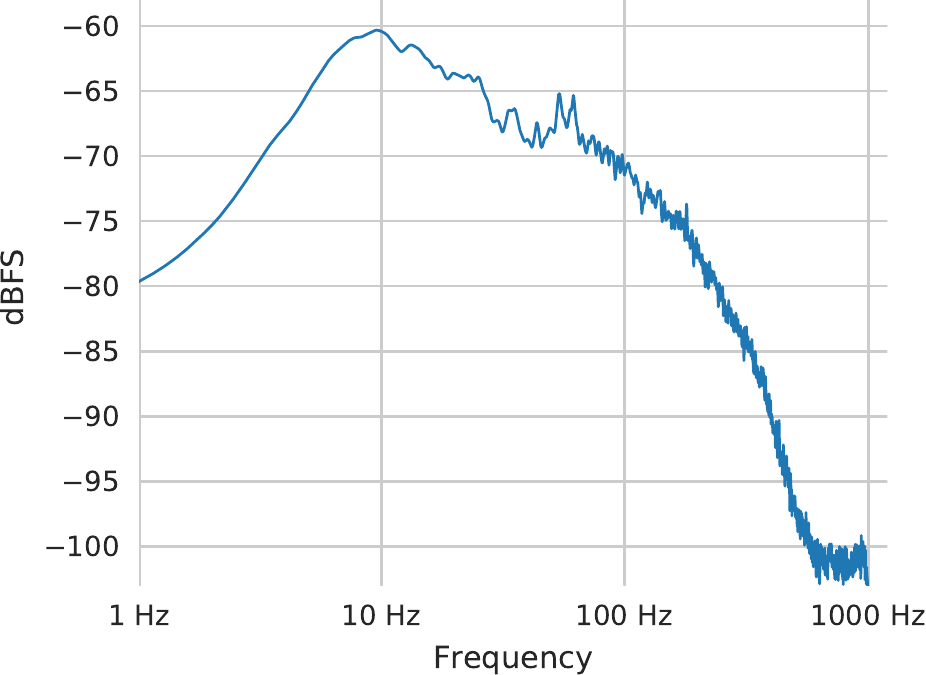}
	\caption[asdf]{
		The average magnitude of the ambient noise level at the test site, in \gls{dbfs}, where \num{0} \gls{dbfs} corresponds to the clipping point of the sound recorder's analog\hyp to\hyp digital converter.
	}
	\label{fig:compare_bg_noise}
\end{figure}

Tests were performed in a quiet location near Pleateu Road in the Cape Peninsula, South Africa.
The ambient noise level at the test site as recorded by our sound recorders is shown in Figure~\ref{fig:compare_bg_noise}.
This was done to ensure that the test environment contains only one sound source, with a good signal\hyp to\hyp noise ratio.

Two specially built infrasonic sound recorders with integrated \gls{gps} modules were placed \SI{147.7}{\meter} apart.
Each recorder has a planar array of \num{6} omni\hyp directional ICS-40300 \gls{mems} microphones placed in a circular pattern with \SI{20}{\centi\meter} radius.

A loudspeaker playing recordings of elephant rumbles and an attached \gls{gps} device provided the sound source and \gls{ssl} ground truth.
The loudspeaker and \gls{gps} device were carried along a meandering path between the recording devices for about \num{4} minutes.

The incoming audio was split into overlapping frames of \SI{1}{\second} each, with a frame skip of \SI{200}{\milli\second}.
For each frame, \glspl{tdoa} between the microphone pairs were estimated using cross correlation and the \gls{doa} for each microphone pair was calculated using Equation~\ref{eq:pair_doa}.
The algorithm proposed in \S\ref{sec:extend} was used with \num{512} histogram bins and a kernel concentration of $\kappa=10$ to estimate the \gls{doa} of the sound source with respect to the microphone array for each window.

\begin{figure*}[h]
	\centering
	\includegraphics[scale=.9]{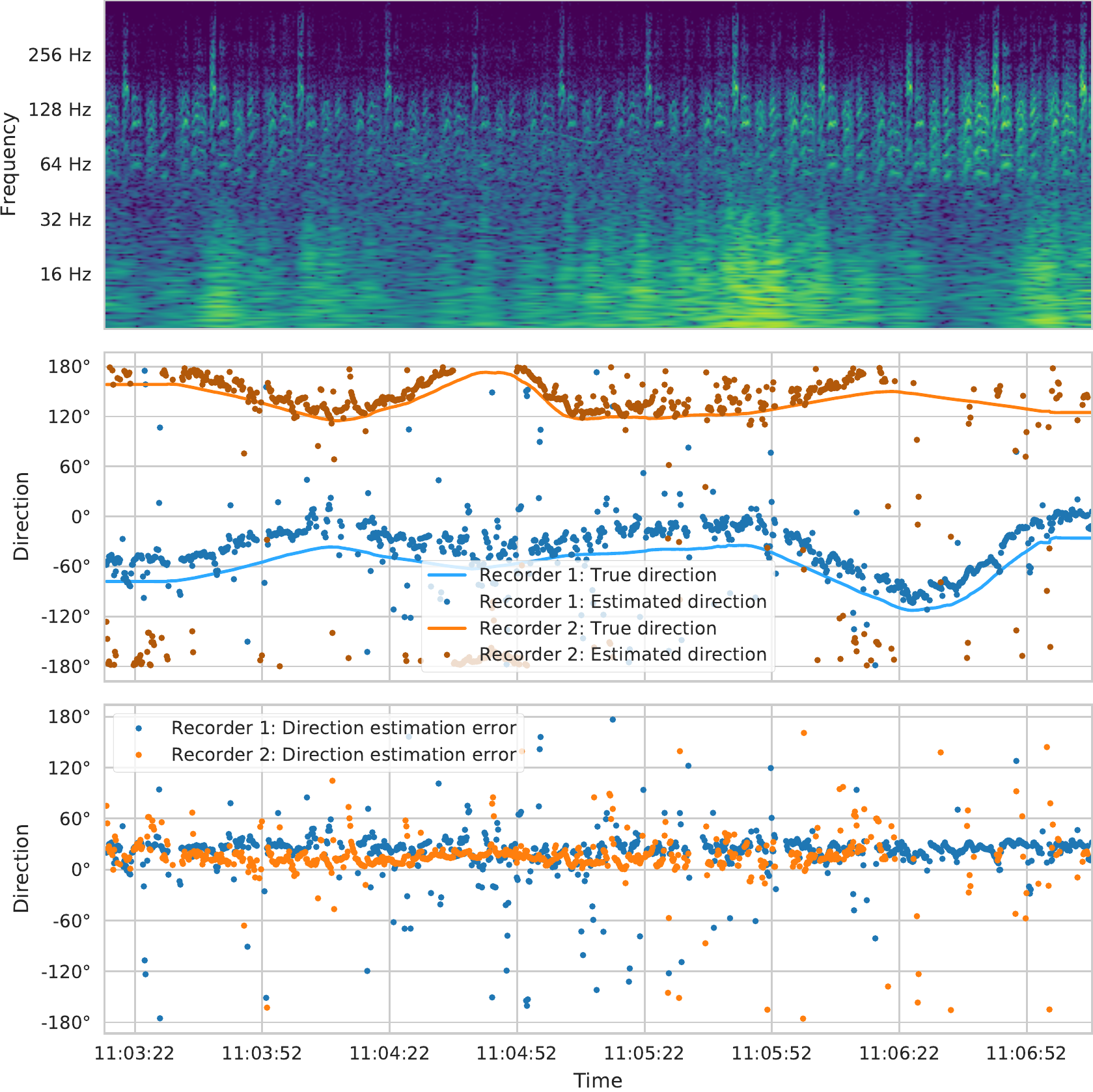}
	\caption[asdf]{
		From top to bottom:
		\begin{enumerate*}
			\item The spectrogram of the recording,
			\item the true \gls{doa} (lines) and estimated \gls{doa} (dots) of the sound sources, and
			\item the errors in the \gls{doa} estimates
		\end{enumerate*}.
		The time axis is shared.
	}
	\label{fig:maan_wyd_rumble_doa_errors_vs_time}
\end{figure*}

Figure~\ref{fig:maan_wyd_rumble_doa_errors_vs_time} shows, from top to bottom:
\begin{enumerate*}
	\item The spectrogram of the recording,
	\item the true (lines) and estimated (dots) \gls{doa} of the sound sources, and
	\item the errors in the \gls{doa} estimates
\end{enumerate*}.
The time axis is shared.
The repeated harmonic patterns seen in the \SI{40}{\hertz} to \SI{256}{\hertz} band are the reproduced elephant rumbles.
Some unwanted low\hyp frequency noise can be seen in the \SI{8}{\hertz} to \SI{32}{\hertz} band, especially at 11:05:52.
The estimated directions deviate from the true directions (calculated using the \gls{gps} coordinates) by a constant value, but in general follow the \gls{ssl} faithfully.

With multiple recorders, these \gls{doa} estimates may be used to estimate the \gls{ssl} as shown in \S\ref{sec:triangulation}, and techniques like Kalman filtering or \gls{phd} filtering may be used to combine these estimates into a single track \cite{hunter2012data}.

\begin{figure*}
	\centering
	\includegraphics[scale=.9]{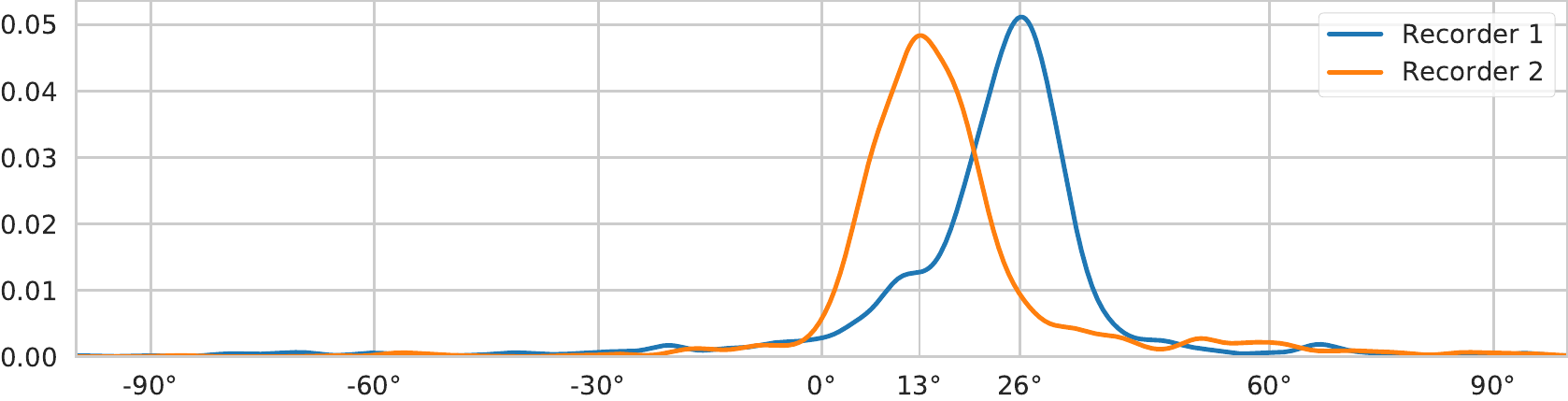}
	\caption{
		Estimation of the distribution of \gls{doa} errors for each recorder.
		The mode (position of the peak) of each distribution is probably equal to the orientation error of the recorder, since each recorder was placed and oriented manually.
		In practice, the \gls{doa} estimates may be improved by subtracting the mode.
		Recorders 1 and 2 provided estimates with standard deviations of \ang{20.4} and \ang{15.2}, respectively.
	}
	\label{fig:maan_wyd_rumble_doa_errors_kde}
\end{figure*}

Figure~\ref{fig:maan_wyd_rumble_doa_errors_kde} shows the estimated distribution of \gls{doa} errors for each recorder.
The standard deviations of these distributions quantify the accuracy of our \gls{doa} estimates.
They are \ang{20.4} and \ang{15.2}, respectively.

The non\hyp zero modes of these errors might be ascribed to the inaccurate manual orientation of the devices during testing.
In practice, such errors may be corrected by adding a constant angle to all \gls{doa} estimates before triangulation or by aligning the devices accurately using laser pointers.

\section{CONCLUSION}

A method of obtaining an unambiguous \gls{doa} estimate from a non\hyp linear, planar microphone array was presented.
All $2^N$ possible interpretations of the ambiguous \gls{doa} estimates from all $M\choose 2$ possible microphone pairs within the $M$\hyp element array were considered.
For every interpretation, the consensus \gls{doa} estimate was deduced from the mode of the \gls{kde} of all \gls{doa} estimates from the $N$ microphone pairs.
The \gls{doa} consensus estimate with the lowest average absolute deviation from the \gls{doa} estimates obtained from the $N$ microphone pairs was chosen as the final \gls{doa}, thereby eliminating the ambiguities that existed in the input data.
Efficient numerical implementation of the algorithm using the \gls{fft} is possible.
Simulations showed a 3:1 reduction in input noise, and practical outdoor tests with two real sound recorders provided \gls{doa} estimates with standard deviations of \ang{20.4} and \ang{15.2}, respectively.

\bibliographystyle{IEEEtran}
\bibliography{IEEEabrv,bib}

\end{document}

%% file: acronyms.tex
\newacronym[
	plural={DOAs},
	firstplural={Directions of Arrival (DOAs)},
]{doa}{DOA}{Direction of Arrival}
\newacronym[
	plural={AOAs},
	firstplural={Angles of Arrival},
]{aoa}{AOA}{Angle of Arrival}
\newacronym[
	plural={TDEs},
	firstplural={Time Delay Estimations (TDEs)},
]{tde}{TDE}{Time Delay Estimation}
\newacronym[
	plural={TDOAs},
	firstplural={Time Differences of Arrival (TDOAs)},
]{tdoa}{TDOA}{Time Difference of Arrival}
\newacronym[
	plural={SSLs},
	firstplural={Sound Source Locations (SSLs)},
]{ssl}{SSL}{Sound Source Location}
\newacronym[
	plural={KDEs},
	firstplural={Kernel Density Estimations (KDEs)},
]{kde}{KDE}{Kernel Density Estimation}
\newacronym[
	plural={RPIs},
	firstplural={Raspberry PIs (RPIs)},
]{rpi}{RPI}{Raspberry PI}
\newacronym{dbfs}{dBFS}{deciBel Full Scale}
\newacronym{gps}{GPS}{Global Positioning System}
\newacronym{utm}{UTM}{Universal Transverse Mercator}
\newacronym{mems}{MEMS}{Microelectromechanical System}
\newacronym{ncg}{NCG}{Newton Conjugate Gradient}
\newacronym[
	plural={FFTs},
	firstplural={Fast Fourier Transforms (FFTs)},
]{fft}{FFT}{Fast Fourier Transform}
\newacronym[
	plural={DFTs},
	firstplural={Discrete Fourier Transforms (DFTs)},
]{dft}{DFT}{Discrete Fourier Transform}
\newacronym{phd}{PHD}{Probability Hypothesis Density}